\documentstyle[12pt,aaspp4]{article}

\slugcomment{Accepted for Astrophys.\ J.\ Letters}

\newcommand{\Casa}{Cas~A}
\newcommand{\Cass}{Cassiopeia~A}

\begin{document}

\title{Evidence of X-ray Synchrotron Emission from Electrons \\
    Accelerated to 40~TeV in the Supernova Remnant \Cass}

\author{G.\ E.\ Allen\altaffilmark{1}, J.\ W.\ Keohane\altaffilmark{2},
    E.\ V.\ Gotthelf\altaffilmark{3}, R.\ Petre, and K.\ Jahoda}
\affil{NASA Goddard Space Flight Center, Laboratory for High Energy
    Astrophysics, Code 662, Greenbelt, MD 20771}

\author{R.\ E.\ Rothschild, R.\ E.\ Lingenfelter, W.\ A.\ Heindl, D.\ Marsden,
    D.\ E.\ Gruber, M.\ R.\ Pelling, and P.\ R.\ Blanco}
\affil{University of California, San Diego, Center for Astrophysics
    \& Space Sciences, Code 0111, 9500 Gilman Drive, La Jolla, CA 92093}

\altaffiltext{1}{ NRC-NASA GSFC Postdoctoral Research Associate;
    glenn.allen@gsfc.nasa.gov}
\altaffiltext{2}{University of Minnesota, Astronomy Department} 
\altaffiltext{3}{Universities Space Research Association}

\begin{abstract}
We present the 2--60~keV spectrum of the supernova remnant \Cass\ measured 
using the Proportional Counter Array and the High Energy X-ray Timing 
Experiment on the Rossi X-ray Timing Explorer satellite.
In addition to the previously reported strong emission-line features produced 
by thermal plasmas, the broad-band spectrum has a high-energy ``tail'' that 
extends to energies at least as high as 120~keV.
This tail may be described by a broken power law that has photon indices of 
$\Gamma_{1} = 1.8^{+0.5}_{-0.6}$ and $\Gamma_{2} = 3.04^{+0.15}_{-0.13}$ and 
a break energy of $E_{\rm b} = 15.9^{+0.3}_{-0.4}$~keV.
We argue that the high-energy component, which dominates the spectrum above
about 10~keV, is produced by synchrotron radiation from electrons that have 
energies up to at least 40~TeV.
This conclusion supports the hypothesis that Galactic cosmic rays are 
accelerated predominantly in supernova remnants.  
\end{abstract}


\section{Introduction}
\label{intro}

Galactic cosmic rays that have energies below $\sim 10^{14}$~eV are believed to
be accelerated predominantly in the shocks of supernova remnants (SNRs).
Prior to the last two years, little experimental evidence existed to support 
this hypothesis.
Radio measurements show that many SNRs emit synchrotron radiation from 
nonthermal electrons, but these measurements are not sensitive to emission 
from electrons that have energies greater than $\sim 10^{10}$~eV.
Gamma-ray measurements at energies $< 10^{10}$~eV also show that some SNRs 
contain nonthermal particles (\cite{esp96}), but the interpretation of these 
results is controversial (\cite{all95}; \cite{les95}; \cite{dej97}; 
\cite{gai97}; \cite{stu97}).
However, recent measurements of X-ray spectra of SNRs have provided the best 
evidence to date that SNRs may be responsible for most of the cosmic-ray 
acceleration in the Galaxy.

While the X-ray spectra of SNRs have traditionally been modeled as thermal 
emission from shock-heated plasmas, the spectra of SN~1006 (\cite{koy95}) and 
RX~J1713.7$-$3946 (\cite{koy97}) are dominated by nonthermal emission 
interpreted to be synchrotron radiation from electrons having energies up to 
$\sim 10^{14}$~eV.  
IC~443 may also contain a site that emits X-ray synchrotron radiation from 
electrons having energies at least as high as $2 \times 10^{13}$~eV 
(\cite{keo97}).

We present the 2--60~keV spectrum of \Cass\ (\Casa) measured using detectors 
on the Rossi X-ray Timing Explorer (RXTE) satellite.  
This spectrum has a high-energy ``tail'' that extends to energies at least as
high as 60~keV.
The strong emission lines in the 0.5--10~keV spectrum (\cite{hol94}; 
\cite{jan88}; \cite{tsu86}) can be described by two-temperature thermal 
plasma models whose flux is dominated by emission from either swept-up material 
(\cite{bor96}) or shocked ejecta (\cite{vin96}; \cite{jan88}).
However, these models do not describe the higher-energy tail of the spectrum
(\cite{fav97}; \cite{pra79}; \cite{hat92}) that has been detected up to at 
least 120~keV (\cite{the96}).
The results of our analysis imply that both the X-ray tail and the radio 
spectrum ($S_{\nu} = 2723 (\nu /1\;\;{\rm GHz})^{-0.77}$~Jy, \cite{baa77}) 
are produced by synchrotron radiation from a common population of nonthermal 
electrons whose spectrum extends up to at least $4 \times 10^{13}$~eV.

\section{Data and Analysis}
\label{data}

Between 1996 March~31 and 1996 April~17, \Casa\ was observed for 186~ks
using the Proportional Counter Array (PCA) and the High Energy X-ray Timing 
Experiment (HEXTE) on the RXTE satellite.
The PCA (\cite{jah96}) and HEXTE (\cite{gru96}) comprise a 2-250~keV 
spectrophotometer that has an energy resolution $\sim 10$--30\% and a 
collecting area of about 7000~cm$^{2}$ at 6~keV and of 800~cm$^{2}$ at 60~keV.

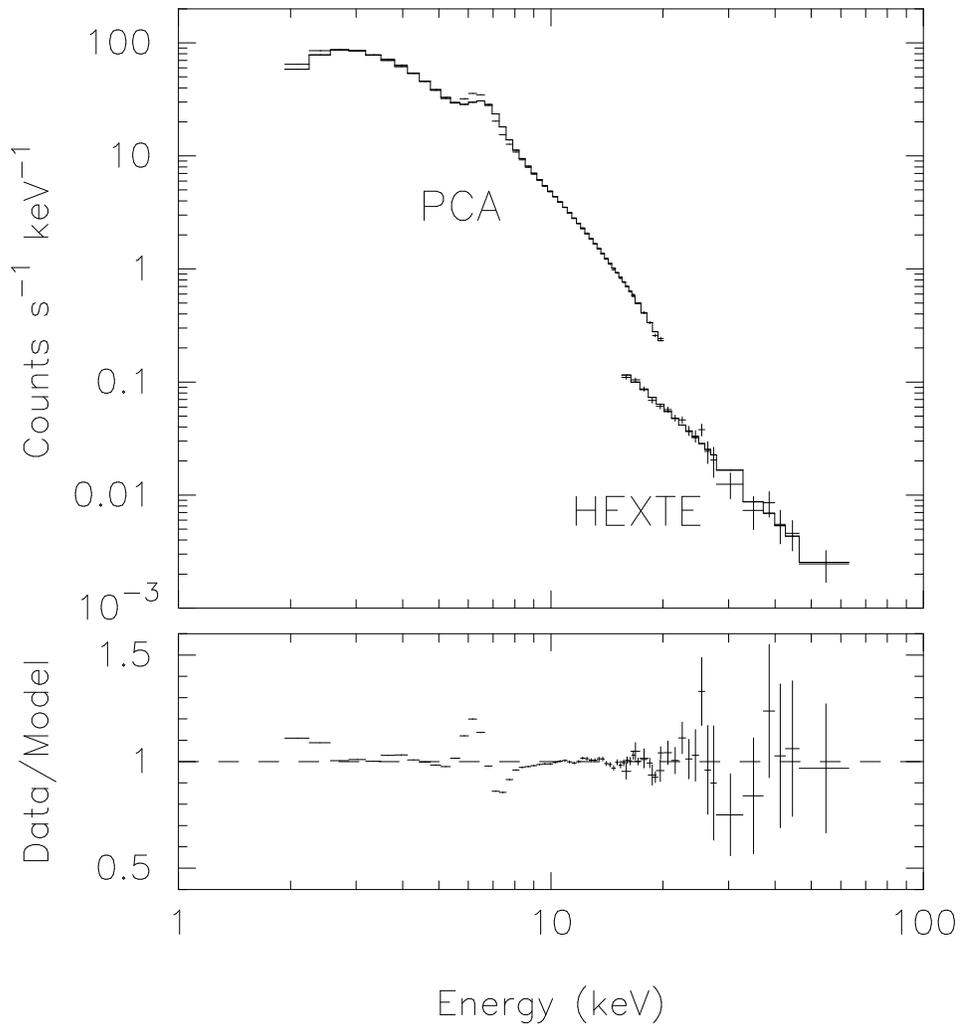
\begin{figure}
\caption{The PCA and HEXTE count spectra of \Casa.
The histogram is the broken power-law model of table~\ref{table1}.
The lower panel shows the ratio of the data to the histogram.
See text for details.
\label{fig1}}
\end{figure}

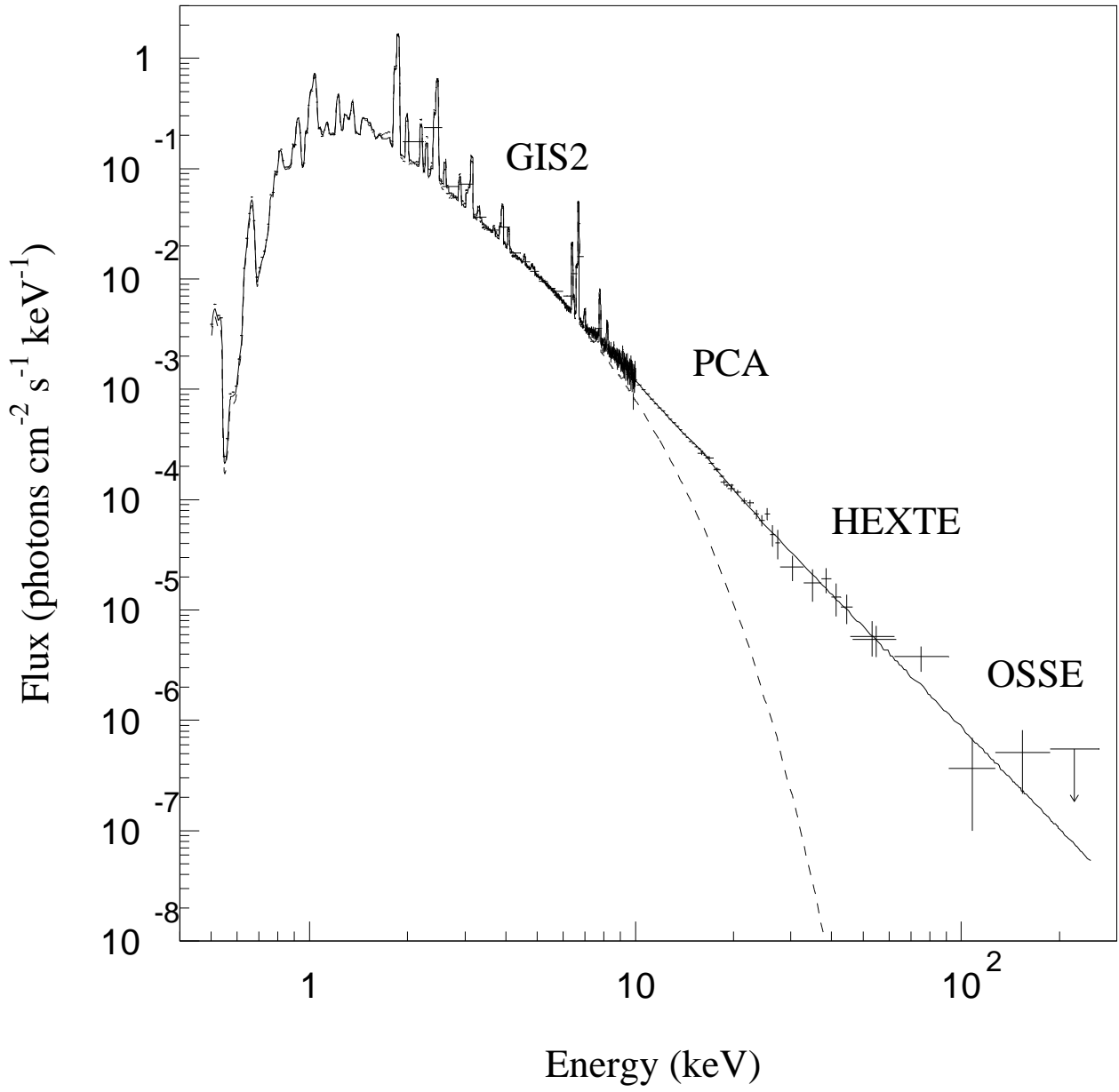
\begin{figure}
\caption{The GIS2, PCA, HEXTE, and OSSE (The et~al.\ 1996) photon spectra of 
\Casa.
The solid curve shows the broken power-law model of table~\ref{table1}.
The dashed curve shows what the shape of the spectrum is if no high-energy
component is included in the fit.
\label{fig2}}
\end{figure}

Figures~\ref{fig1} and~\ref{fig2} show the PCA and HEXTE spectra of \Casa\ 
for 91~ks of the PCA data and for 96~ks of HEXTE data.
Because the strong emission lines may be fit by models with two thermal 
plasmas and with fluorescent K$\alpha$ emission from iron in dust grains 
(\cite{bor97}), we model these features using two Raymond-Smith components 
and a 6.40~keV Gaussian.
While such a model provides an adequate fit to the 0.5--10~keV spectrum, it 
does not fit the spectrum above $\sim 10$~keV (fig.~\ref{fig2}).
At least one more component is needed to model the broad-band X-ray spectrum.

We fit the broad-band spectrum using models that include only one additional 
component that is either a thermal bremsstrahlung, a power law, a power law 
that has an exponential cutoff, or a broken power law.  
In general, the high-energy spectrum is better fit by the nonthermal 
components than a thermal bremsstrahlung component.
Furthermore, the spectrum is better fit by a broken power law than a power law
or a cut off power law.
Table~\ref{table1} summarizes the best fit models that include either a 
broken power law component or a thermal bremsstrahlung component.
Abundances of some line-emitting elements are listed in the table because the 
abundances depend on the high-energy component.
The abundances should be regarded as rough estimates because a better 
treatment requires plasmas that are not in ionization equilibrium.
The values of $\chi^{2}$ in table~\ref{table1} were computed for only the
$\ge 10$~keV portion of the PCA and HEXTE spectra where the high-energy 
component dominates the flux.
The ``$1 \sigma$'' errors quoted in this paper, which may be underestimated, 
were computed by finding the extrema of the $\chi^{2}$ contour that has a 
value of $\chi^{2}$ which is larger than the minimum value by 2.3.

Figure~\ref{fig1} shows the spectrum of the best fit model ``folded'' through 
the PCA and HEXTE response matrices.
The lower panel of this figure shows the ratio of the measured spectra to 
this model.
Aside from the uncertainty in the cross calibration of the PCA and HEXTE, this
ratio is dominated by systematic errors in the response matrix and the
background spectrum for the PCA and by statistical errors for the HEXTE.
For example, the features near 2~keV and between 5 and 8~keV in the lower 
panel of figure~\ref{fig1} reflect inaccuracies of the v2.0.2 PCA response 
matrix that are observed when modeling PCA spectra for other sources, such as 
the Crab (\cite{jah96}).
These features are not evident in the fit to the spectrum of the Gas Imaging
Spectrometer no.\ 2 (GIS2) on the Advanced Satellite for Cosmology and
Astrophysics.

Figure~\ref{fig2} shows the GIS2, PCA, HEXTE, and OSSE (Oriented Scintillation
Spectrometer Experiment on the Compton Gamma Ray Observatory satellite, 
\cite{the96}) spectral data and the ``unfolded'' spectrum of the best fit 
model (solid curve).
For comparison, figure~\ref{fig2} also shows the spectrum of a model that
includes only the two Raymond-Smith components and the Fe~K$\alpha$ dust
component (dashed curve).

\section{Discussion}
\label{disc}

The spectra of several emission mechanisms---thermal and nonthermal 
bremsstrahlung, a pulsar, inverse Compton scattering, and synchrotron 
radiation---could be described by a broken power law in the range of
10--63~keV.
Each possibility is discussed in the context of \Casa.

It is unlikely that the high-energy X-ray tail of \Casa\ is thermal.
As shown in figure~\ref{fig2} and table~\ref{table1}, the broad-band X-ray
spectrum is not well fit by two or three thermal component models.
Furthermore, a physical interpretation of a model that has three thermal
components is difficult.
Two of the components could be associated with the forward-shocked 
circumstellar material and the reverse-shocked ejecta.
The third might result from a nonuniform distribution of the circumstellar or 
ejected matter.
However, spatially-resolved spectroscopy shows that \Casa\ has similar plasma 
conditions across the remnant (\cite{hol94}).

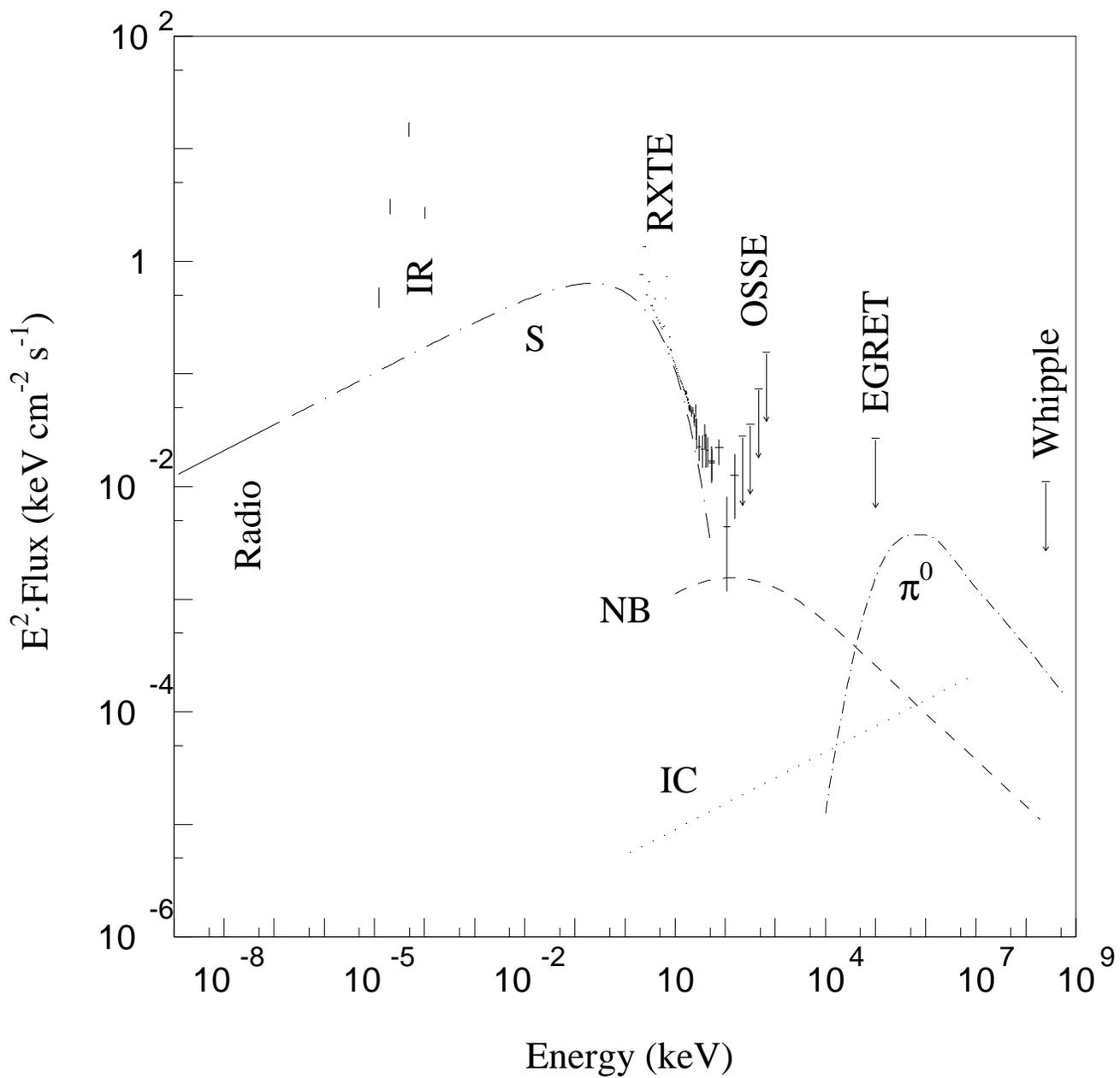
\begin{figure}
\caption{The multiwavelength photon spectrum of \Casa.
The four broken curves are estimates of the fluxes from synchrotron radiation
(S), from nonthermal bremsstrahlung (NB), from inverse Compton scattering of 
the cosmic microwave background (IC), and from the decay of neutral pions 
($\pi^{0}$).
See text for details.
\label{fig3}}
\end{figure}

Asvarov et~al.\ (1989) suggest that the high-energy X-ray spectrum of \Casa\ 
is dominated by nonthermal bremsstrahlung emission.
However, an estimate of the nonthermal bremsstrahlung spectrum 
(fig.~\ref{fig3}) yields a 20--50~keV effective photon index of $\sim 1.9$,
which is significantly smaller than the fitted index of $3.04^{+0.15}_{-0.13}$.
Therefore, the X-ray tail of \Casa\ appears to be too steep to be consistent 
with a nonthermal bremsstrahlung emission process.

Although pulsars can produce nonthermal X-ray continua, it is unlikely that the
high-energy tail of \Casa\ is produced by a pulsar.
Analyses of X-ray (\cite{all97}; \cite{pra79}), radio (\cite{woa93}) and 
optical (\cite{hor71}) data reveal no evidence of pulsation and no bright 
point-like feature is observed near the center of the remnant (\cite{fab80}; 
\cite{hol94}; \cite{jan88}).

An inverse Compton spectrum could be consistent with the shape of the tail, but
the estimated flux is a factor of $\sim 10^{4}$ smaller than the measured 
flux at energies in the range of 10--63~keV (fig.~\ref{fig3}).

In contrast to the other emission processes, both the shape and the flux of 
the high-energy component of the X-ray spectrum of \Casa\ are consistent with 
a synchrotron radiation mechanism.
In general, the synchrotron spectrum of a SNR is expected to span the entire 
range from radio to X-ray energies and to steepen gradually at or near X-ray
energies (\cite{rey97}).
This expectation is consistent with the multiwavelength spectrum of \Casa\ 
because the extrapolation of the radio synchrotron spectrum to X-ray energies 
does not lie above the observed thermal emission at infrared wavelengths or
below the X-ray spectrum (fig.~\ref{fig3}) and because the broken power law of
table~\ref{table1} approximately describes a gradually steepening spectrum.
The synchrotron spectrum of figure~\ref{fig3} is a power-law spectrum with a 
$e^{- \sqrt{E_{\gamma}/1~{\rm keV}}}$ cutoff.
This photon spectrum corresponds to a power-law electron spectrum with a 
$e^{- (E_{\rm e}/8\times10^{12}~{\rm eV})}$ cutoff and assumes that the
spectral distribution of synchrotron photons emitted by electrons of a common
energy is a delta function. 
A better estimate of the synchrotron spectrum, which is expected to steepen
more gradually than $e^{- \sqrt{E_{\gamma}/1~{\rm keV}}}$ (\cite{rey97}), 
requires a numerical simulation of the particle acceleration conditions 
appropriate for \Casa\ including the effects (1) of synchrotron losses on the 
electron spectrum, (2) of possible curvature in the electron spectrum 
(\cite{mez86}; \cite{ell91}), and (3) of the spectral distribution of 
synchrotron photons for electrons of a given energy.

Figure~\ref{fig3} shows the radio (\cite{baa77}), infrared (\cite{mez86}), 
RXTE, OSSE (\cite{the96}), EGRET (\cite{esp96}), and Whipple (\cite{les95}) 
spectral data of \Casa.
The estimated synchrotron spectrum is a single power law that has a
$e^{-\sqrt{E_{\gamma}/1~{\rm keV}}}$ cutoff.
Also shown are estimates of the photon spectra produced by nonthermal
bremsstrahlung, by inverse Compton scattering of the cosmic microwave 
background radiation, and by the decay of neutral pions.
These latter three estimates were computed for electron and proton spectra 
that have a common spectral index of 2.54 using equation~13 and figure~3 of 
Gaisser et~al.\ (1997).  
The electron density spectrum is assumed to be 
$7 \times 10^{-8}$~cm$^{-3}$~GeV$^{-1}$ at 1~GeV.
The density of accelerated nuclei is assumed to be one hundred times larger 
than the density of accelerated electrons independent of energy.
The nonthermal bremsstrahlung and $\pi^{0}$ decay spectra were computed for a
density of nuclei in \Casa\ of 30~cm$^{-3}$.
The shape of the nonthermal bremsstrahlung spectrum at energies below 1~MeV is
based on the shape of the nonthermal bremsstrahlung spectrum of Sturner
et~al.\ (1997, fig.~11).

\section{Conclusion}
\label{sum}

The RXTE data reveal that the X-ray continuum of \Casa\ has a nonthermal 
high-energy ``tail.''
The tail (1) is qualitatively consistent with a simple model of synchrotron 
emission from SNRs, (2) is inconsistent with the expected shape of a nonthermal 
bremsstrahlung spectrum, and (3) is inconsistent with the estimated inverse 
Compton flux.
No evidence of pulsation or of a bright central point-like feature has been
reported.
Therefore, the tail is most likely produced by synchrotron radiation.

X-ray synchrotron emission has important implications for particle 
acceleration in \Casa.
The energy of a synchrotron photon, $E_{\gamma}$, is related to the energy of 
the emitting electron and the magnetic field strength by
$E_{\gamma} \sim 0.6 B_{\mu \rm G} E_{14}^{2}$~keV, where $B_{\mu \rm G}$ is
the magnetic field strength in units of $\mu$G and $E_{14}$ is the energy of 
the electron in units of $10^{14}$~eV.
If the synchrotron spectrum extends to energies at least as high as 120~keV and
the magnetic field is 1~mG (the ``equipartition'' value, e.g.\ \cite{lon94}), 
the accelerated electron spectrum extends up to at least $4 \times 10^{13}$~eV.
Since electrons and protons are expected to be accelerated in the same manner 
at energies much larger than the rest mass energy of the proton 
(\cite{ell91}), this result provides strong evidence for the acceleration of 
cosmic-ray protons in \Casa\ to the same energies.
The calculation of the equipartition value of the magnetic field leads to an
estimate of the total amount of energy in accelerated protons and electrons of
$\sim 3 \times 10^{49}$~erg (\cite{lon94}).
This amount of energy is comparable to the average amount of energy required 
from a Galactic SNR ($\sim 3$--$10 \times 10^{49}$~erg), over the lifetime of 
the remnant, if all of the Galactic cosmic rays are accelerated in SNRs.

\Casa\ is the fourth SNR reported to exhibit evidence of X-ray synchrotron
radiation.
Collectively, the results for \Casa, SN~1006 (\cite{koy95}), RX~J1713.7$-$3946
(\cite{koy97}), and IC~443 (\cite{keo97}) support the hypothesis that 
Galactic cosmic rays are accelerated predominantly in SNRs.

\acknowledgments

We thank Apostolos Mastichiadis and Steve Reynolds for advice about the
expected shape of the spectrum of synchrotron radiation.
We are grateful to Jacco Vink for many stimulating and thoughtful discussions 
about the data and about the implications of the data.
We appreciate Jennifer Allen's careful review of the manuscript.
This work was performed while GEA held a NRC-NASA GSFC Research Associateship.
The work at UCSD was supported by NASA contract NAS5-30720 and NASA grant
NAG5-3375.


\clearpage

\clearpage
 
\begin{deluxetable}{cccccccccc}
\footnotesize
\tablecaption{Parameters of the Spectral Models\tablenotemark{a} \label{table1}}
\tablewidth{0pt}
\tablehead{ \multicolumn{10}{c}{{Broken Power Law}} \\ 
    \cline{1-10}
    & & & & \multicolumn{5}{c}{{Abundance}\tablenotemark{b}} & \\ 
    \cline{5-9}
    $\Gamma_{1}$ & $E_{\rm b}$ & $\Gamma_{2}$ & Flux at 1~keV & Si & S & Ar 
    & Ca & Fe & 
    $\chi^{2}$\/\tablenotemark{c} \\
    $(E<E_{\rm b})$ & (keV) & $(E>E_{\rm b})$ 
    & (ph~cm$^{-2}$~s$^{-1}$~keV$^{-1}$) & \multicolumn{5}{c}{(Solar)} 
    & (69 dof)
}
\startdata
$1.8^{+0.5}_{-0.6}$ & $15.9^{+0.3}_{-0.4}$ & $3.04^{+0.15}_{-0.13}$ & 0.038  
    & 2.1 & 4.3 & 8.3 & 2.6 & 0.6 & 105 \\ \hline \hline
\multicolumn{10}{c}{Thermal Bremsstrahlung} \\ \cline{1-10} 
\multicolumn{3}{c}{$kT$} 
    & \multicolumn{1}{c}{{$\int n_{\rm e} n_{\rm i} dV$}\tablenotemark{d}}
    & \multicolumn{5}{c}{{Abundance}\tablenotemark{b}} & 
    $\chi^{2}$\/\tablenotemark{c} \\ \cline{5-9}
\multicolumn{3}{c}{(keV)} & \multicolumn{1}{c}{($10^{58}$ cm$^{-3}$)}
    & \multicolumn{5}{c}{(as above)} & (71 dof) \\ \hline
\multicolumn{3}{c}{$23.7^{+0.8}_{-0.7}$} 
    & \multicolumn{1}{c}{{$1.16 \pm 0.04$}}
    & 2.0 & 4.1 & 7.7 & 2.3 & 0.6 & 236 \\
\enddata

\tablenotetext{a}{These fits also include two Raymond-Smith components that 
    have temperatures of $kT_{1} = 0.6$~keV and $kT_{2} = 2.9$~keV and a 
    6.40~keV Gaussian component ($1.3 \times 10^{-3}$
    photons~cm$^{-2}~$s$^{-1}$) for K$\alpha$ emission from iron in dust
    grains.
    The fit column density $n_{\rm H} = 1.1 \times 10^{22}$ H~atoms~cm$^{-2}$.}
\tablenotetext{b}{The Si, S, and Ar abundances pertain to the 
    $kT_{1} = 0.6$~keV plasma.  The Ca and Fe abundances pertain to the 
    $kT_{2} = 2.9$~keV plasma.}
\tablenotetext{c}{For the energy range from 10 to 63~keV.}
\tablenotetext{d}{The quantities $n_{\rm e}$ and $n_{\rm i}$ are the electron 
    and ion densities respectively.  
    The average value of $n_{\rm e} n_{\rm i}$ is $\sim 25$~cm$^{-6}$ for a 
    distance of 3.4~kpc and a filling fraction of 1/4.} 

\end{deluxetable}

\end{document}